\documentclass{ws-procs9x6}

\begin{document}

\title{Exclusive $\rho^0$ electroproduction on the proton :\\ 
GPDs or not GPDs ?}

\author{M. Guidal$^*$}

\address{Institut de Physique Nucl\'eaire d'Orsay \\
91405 Orsay, FRANCE\\ 
E-mail: guidal@ipno.in2p3.fr}

\author{S. Morrow}

\address{Institut de Physique Nucl\'eaire d'Orsay \\
91405 Orsay, FRANCE\\
\& CEA-Saclay, Service de Physique Nucl\'eaire\\ 
91191 Gif- sur-Yvette,Cedex, FRANCE}

\begin{abstract}
We discuss the interpretation of the $ep\to ep\rho^0$ process in terms of,
on the one hand, Generalized Parton Distributions and, on the other hand, 
an effective hadronic model based on Regge theory.
\end{abstract}


\bodymatter

\vskip 1.5cm
The $\gamma^{(*)}p\to p\rho^0$ process is, due to its relatively high cross 
section over a wide range of energy (from threshold at $W\approx$ 1.75 GeV 
up to $W\approx$ 200 GeV, where $W$ is the center of mass energy of the $(\gamma,p)$ 
system), one of the exclusive processes on the proton the most studied experimentally
and, consequently, also theoretically.

\begin{figure}[hbtf]
\begin{center}
\psfig{file=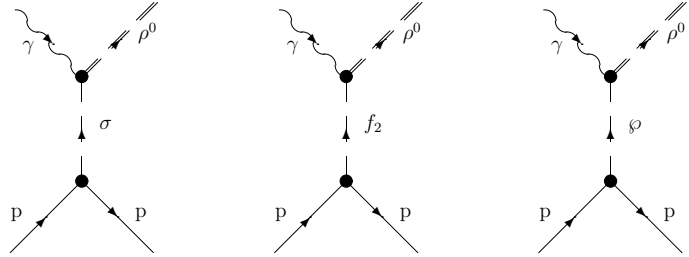,width=4.5in}
\end{center}
\vskip-2.cm
\caption{The dominant $t$-channel meson exchange diagrams for the reaction 
$\gamma p \rightarrow p\rho^0$.}
\label{fig:tregge}
\end{figure}

In photoproduction above the nucleon resonance region 
the reaction is understood to proceed at low $W$ through the exchange of 
mesons ($\sigma$, $f_2$,...) in the so-called $t$-channel (in the form of Regge 
trajectories) and at high $W$ through the exchange of the Pomeron 
trajectory, which carries the quantum numbers of the vacuum (see Fig.~\ref{fig:tregge}).
These diagrams are based on {\it  hadronic} degrees of freedom and the
complex physics of QCD (Quantum Chromo-Dynamics) 
is absorbed into effective meson-nucleon and photon-meson 
coupling constants plus potentially additionnal hadronic form factors.

\begin{figure}[hbtf]
\begin{center}
\psfig{file=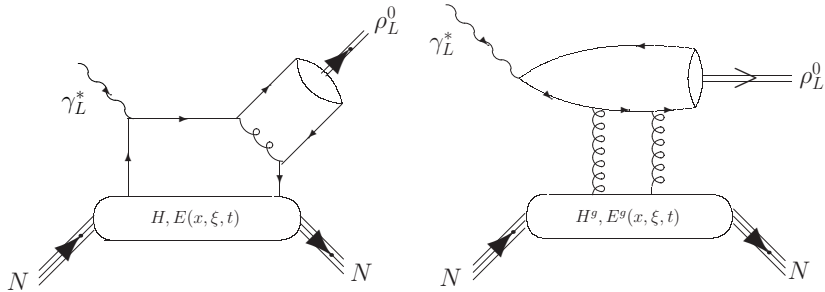,width=4.5in}
\end{center}
\vskip-2.cm
\caption{The ``handbag" diagrams for (longitudinal) vector meson leptoproduction.
``Quark" GPDs are at play on the left and ``gluon" GPDs on the right.}
\label{fig:qandg}
\end{figure}

Going to electroproduction, the increasing virtuality $Q^2$
of the initial photon allows us to probe shorter and shorter distances, 
and be sensitive to {\it partonic} (quarks and gluons) degrees of freedom. 
Then, at sufficiently high $Q^2$, the process should be understandable
in terms of the so-called ``handbag" diagrams illustrated in Fig.~\ref{fig:qandg},
where the incoming virtual photon scatters directly off a quark
whose interaction can be calculable in the framework of perturbative
QCD. The complex non-perturbative QCD structure of the nucleon is then
represented by (quark and gluon) Generalized Parton Distributions (GPDs),
which were shown to factorize from the QCD perturbative process
in the handbag process for incoming \underline{longitudinal} 
photons~\cite{Collins97}.

The GPDs might actually provide the most complete information accessible 
on the structure of the nucleon because they describe the (correlated) spatial and momentum 
distributions of the quarks in the nucleon (including the polarization aspects), its 
quark-antiquark content, a way to access the orbital momentum of the quarks, etc.
We refer the reader to Refs~\cite{goeke,revdiehl,revrady}, for example, for 
very detailed and quasi-exhaustive reviews on the GPD formalism and the definitions
of some of the variables that will be employed in the following.

The question then arises for what ranges in $Q^2$ are the hadronic and partonic
descriptions valid and applicable. A first very simple image that one can form 
is that the {\it quark} handbag diagram (Fig.~\ref{fig:qandg} left) might be associated with 
the meson Regge exchange mechanism (Fig.~\ref{fig:tregge} left and center) while 
the {\it gluon} handbag diagram (Fig.~\ref{fig:qandg} right) might correspond to 
the exchange of the quantum numbers of the vacuum (as gluons form an isosinglet state) 
and thus to the Pomeron exchange mechanism. Indeed, ``gluonic" GPDs, which obviously reflect the glue 
content of the nucleon, are expected to contribute at high $W$, i.e. in the low $x_B$
``sea" domain, while the ``quark" GPDs should be predominant at low $W$,
i.e. in the large $x_B$ ``valence" domain, where $x_B$ can be related to $W$ by 
$x_B\approx \xi=\frac{Q^2}{W^2+Q^2-m_N^2}$.

This correspondence is not fully exact in the sense that
the {\it quark} handbag diagram can also carry the quantum numbers of the vacuum
when sea quarks are exchanged and thus contribute to a Pomeron-type mechanism.
Also, GPDs are not only $t$-channel mechanisms (which are just the $|x| < \xi$ part of
the GPDs), there is also a part associated with standard ``diagonal" Parton Distribution 
Functions (PDFs), which is not contained in $t$-channel Regge exchanges.

Experimentally, in the high $W$ domain, data for the exclusive electroproduction of the $\rho^0$ meson
have been obtained at several $Q^2$ values by the H1 and ZEUS (at HERA), 
NMC (at CERN) and E665 (at Fermilab) experiments.
It was shown by Frankfurt et al.~\cite{Fra96} that
these high $W$ data can be well interpreted in terms of the 2-gluon 
exchange mechanism of Fig.~\ref{fig:qandg} right.
However, in order to achieve this, because the $Q^2$ domain
where the data exist is relatively low, corrections have to be incorporated
to the pure leading-twist calculation of the handbag diagrams. The straightforward
leading twist calculation leads at low $Q^2$ values to an overestimate of 
the data by a large factor (up to $2$ at $Q^2$=10 GeV$^2$ and higher
at smaller $Q^2$). Also, the leading-twist calculation predicts
that $\frac{d\sigma_L}{dt}$ should behave as $\frac{1}{Q^6}$, at fixed 
$x_B$ whereas a flatter $Q^2$ dependence is observed in the data. 

The origin of the overestimate of the data lies in the presence of gluon
exchange and the associated strong coupling to quarks $\alpha_s$ 
in the handbag diagrams of Fig.~\ref{fig:qandg}. Indeed, $\alpha_s$
is ``running" and, in particular, is rising as $Q^2$, or more generally,
the scale associated to the quark-gluon vertex (to which $Q^2$
is proportional), decreases. Also, the gluon exchange 
is associated with a propagator of the form $\frac{1}{zQ^2}$, where
$z$ is the momentum fraction carried by the quark(s) interacting
with the gluon. As $z$ runs between $0$ and $1$, 
at low $z$, this can lead to strong enhancements of the cross section.

The handbag diagrams deal mainly with longitudinal degrees of 
freedom ($x,\xi,...$). The idea brought by Frankfurt et al. to
remedy the ``overshooting" of the leading twist calculation
was to also take into account the {\it transverse} 
momentum (``$k_\perp$") dependence in the calculations of the handbag processes. 
This approach, also called the ``modified parturbative approach",
was originally initiated in Ref.~\cite{mpa}. The
consequence is that there will always be a ``minimum" momentum scale 
(the average transverse momentum) and therefore singularities of the type $z\to 0$
will be much avoided. With such a prescription, it was found a remarkable agreement
for the $x_B$ (mainly driven by the momentum fraction distribution
of the gluons) dependence and the $Q^2$ (asymptotic $\frac{1}{Q^6}$
modulated by the ``$k_\perp$" correction factor) dependences of the $ep\to e p\rho^0$
cross section (as well as for the absolute normalization).

We have just discussed the high $W$ domain where, in summary, the $\rho^0$ electroproduction data
look interpretable in terms of the 2-gluon exchange mechanism of Fig.~\ref{fig:qandg} 
right (modulo ``$k_\perp$" corrections). Now, in the low $W$ region, we ask whether, similarly,
the quark handbag diagram of Fig.~\ref{fig:qandg} left (modulo ``$k_\perp$" corrections) is
also at play ? The few data that exist for the $ep\to e p\rho^0$
process at low $W$ include 
early data (1976) with a 7.2 GeV electron beam at DESY~\cite{JoosRho}
and with a 11.5 GeV electron beam at Cornell~\cite{Cassel} (1981) and more
recently, with a 470 GeV muon beam at Fermilab~\cite{e665}, a 27 GeV positron beam 
at HERMES~\cite{HERMESrho}, and a 4.2 GeV electron beam at JLab~\cite{cynthia} using 
the CLAS detector. Currently, JLab experiment E99-105~\cite{E99105} which recently took
data with a 5.75 GeV electron beam and the CLAS detector is in the final stage
of data analysis and we will have here a quick snapshot to some 
\underline{PRELIMINARY} results.

The amplitude for the ``quark" handbag diagram of Fig.~\ref{fig:qandg} left
(with ``$k_\perp$" corrections) has been calculated, for instance, in Ref.~\cite{marcprl1}.
The main input to this calculation is the parametrization of the GPDs.
As a first approximation, only $H(x,\xi,t)$ is considered in the following.
The $(x,\xi)$ dependence of $H$ is parametrized in terms of the Double Distributions 
ansatz~\cite{Mu94,RadyDD} and the $(x,t)$ correlation is based on Regge 
theory~\cite{gprv}~:

\begin{equation}
GPD^q(x,\xi,t)=\int d\alpha d\beta \delta(x-\beta-\xi\alpha)DD(\alpha,\beta,t)
\end{equation}
with~:
\begin{equation}
DD(\alpha,\beta,t)=h(\beta,\alpha)q(\beta)\beta^{-\alpha^\prime (1-\beta) t}
\end{equation}
and 
\begin{equation}
h(\beta,\alpha)=\frac{\Gamma(2b+2)}{2^{2b+1}\Gamma^2(2b+1)}
\frac{\left[(1-\mid\beta\mid)^2-\alpha^2\right]^b}{(1-\mid\beta\mid)^{2b+1}}
\label{eq:dd}
\end{equation}

\noindent The only free parameters are $\alpha^\prime$ which is fit, and strongly
constrained, to reproduce the nucleon form factors~\cite{gprv} and $b$
which governs the $(x,\xi)$ dependence but has, in effect, relatively
little influence on the cross section.

As a second approximation, in the following, the quark and gluon handbags 
of Fig.~\ref{fig:qandg} are summed at the {\it cross section} level.
In Ref.~\cite{gk}, this sum is treated at the {\it amplitude} level. It 
has been checked that, except in the intermediate $W$ region where
interference is maximal, the two (independent) calculations are in remarkable 
agreement.

\begin{figure}[hbtf]
\begin{center}
\psfig{file=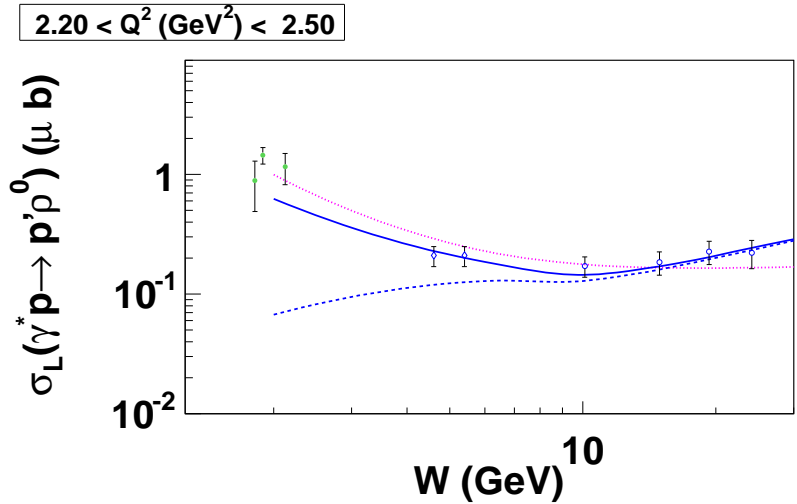,width=4.5in}
\end{center}
\caption{$W$ dependence of the $\gamma_L p\to p\rho^0$ at
$Q^2\approx 2.35$ GeV$^2$. Dashed curve : the ``standard" handbag calculation 
(quark and handbag diagrams added incoherently) based on the Double Distribution
ansatz for the quark GPDs. The thin solid curve is the result of the calculation 
including a (``renormalized") D-term-like contribution.
The dotted curve is the result of the Regge JML calculation.}
\label{fig:w}
\end{figure}

Fig.~\ref{fig:w} shows the total {\it longitudinal} cross section for the exclusive 
$\rho^0$ electroproduction on the proton as a function of $W$ over
a wide range for $Q^2\approx 2.35$
GeV$^2$ with the current world's data. The cross sections exhibit clearly two different 
behaviors as a function of $W$~: starting from low $W$, the cross section decreases 
with $W$. Then, at $W$ around $\approx$ 10 GeV, the $W$ slope changes and the 
cross section slowly rises. The handbag calculation (sum of the quark and gluon 
processes), just described above, shown by the dashed curve, 
gives a decent description of the high and intermediate $W$ region down to $\approx$
6 GeV. This result was already observed by the HERMES 
collaboration~\cite{HERMESrho}. In particular, at high $W$, the rise of the cross 
section is due to the gluon and sea contributions.

Now, focusing on lower $W$ values, the GPD calculation clearly misses
the data. This discrepancy can reach an order of magnitude at the lowest 
$W$ values. The trend of the GPD calculations is to decrease as $W$ 
decreases whereas the data increase. The trend of the GPD calculation
is readily understandable~: GPDs are approximately proportional to 
the forward quark densities $q(x)$ (the relation is not direct
since, among other aspects, the quark densities are, in the Double Distributions,
convoluted with a meson distribution amplitude but, still, the main trends
remain ; in the following argument, we take $x=x_B$). Therefore, as $x$ increases 
($W$ decreases), they tend to $0$
($q(x)\approx (1-x)^3$ for $x$ close to $1$). There might be a slight local increase 
around $x\approx$ 0.3, due to the valence 
contribution (which is slightly apparent on Fig.~\ref{fig:w}) but it can 
never explain the increase of an order of magnitude, especially at low $W$.

Fig.~\ref{fig:w} also shows (dotted curve) the results of the 
calculation based on 
the ``hadronic" $t$-channel diagrams of Fig.~\ref{fig:tregge} which are ``reggeized". 
This is the co-called JML model~\cite{RgModel2} and it reproduces fairly well the two general behaviors 
with $W$ just mentionned. Here, the drop of the cross section at low $W$ is due to 
the $t$-channel $\sigma$ and $f_2$ meson exchange diagrams (the intercept $\alpha(0)$ 
of the $f_2$ trajectory is $\approx 0.5$ and therefore the cross sections decrease with 
energy as $\frac{1}{s^{0.5}}$). The slow rise of the cross 
section which is observed above W $\approx$ 10 GeV is 
attributed to the Pomeron trajectory which has an intercept
$\alpha(0)\approx 1+\epsilon$.

Coming back to the GPD calculation, the conclusions that one can draw are 
two-fold~:
\begin{itemize}
\item The handbag mechanism formalism is not at all the dominant 
mechanism in the low $W$ (valence) region and higher twists or (so far) 
uncontrollable non-perturbative effects suppress the handbag mechanism. 
In this case, one needs to explain why the handbag mechanism works at 
high/intermediate $W$ (low $x_B$) domain and, quite abruptly, no more in 
the valence region. Higher twist can certainly depend on energy but such 
a strong variation with $W$ is certainly puzzling. 
\item An alternative explanation, based on and supported by the success
of the handbag mechanism at high and intermediate $W$ values,
is that the GPD formalism is indeed at work in the valence region but 
that a significant and fundamental contribution, besides the Double Distributions, 
is missing in our parametrization of the GPDs. 
\end{itemize}

What could such a missing contribution in the low ``W" regime be ?
In the framework of the JML model, the strong
rise of the cross section as $W$ decreases is due to the 
$t$-channel $\sigma$ and $f_2$ meson exchange processes.
It could then be tempting to associate the potentially missing piece in the 
GPDs with $t$-channel meson exchanges. The best example of such a contribution
in GPDs is the so-called $D$-term which was originally introduced in 
Ref~\cite{weiss}, where it was shown that it was required to introduce such a term in
the most general parametrisation of GPDs in addition to Double Distributions,
in order to satisfy the polynomiality rule. We recall that the D-term is non-zero
only in the $-\xi < x < \xi$ domain, it is odd in $x$ and it is usually
parametrized in terms of Gegenbauer polynomials of argument $\frac{x}{\xi}$.
The $-\xi < x < \xi$ region corresponds to the $q\bar{q}$ component of the GPDs
and therefore
the D-term can be thought of as representing the exchange of mesonic
degrees of freedom  in the $t$-channel.
We see that a structure that exists only in the $-\xi < x < \xi$ domain would naturally
provide a contribution that decreases with $W$, since $W \approx \frac{1}{\xi}$.
In other words, as $W$ increases $\xi$ decreases, and therefore the support
of such structure decreases and, as a consequence, its contribution diminishes
up to $0$ in the $\xi=0$ limit.

The $D$-term is associated to a Lorentz {\it scalar} in the 
GPD definition and thus accounts only for {\it scalar} (isosinglet) meson exchange 
(as the $\sigma$ meson). Other meson contributions (for instance, tensor as the $f_2$ meson) 
are not included in the $D$-term. However, there is, in principle, in 
the GPD formalism, no reason to restrict these $q\bar{q}$ contributions to 
the scalar component as it is imposed 
in the $D$-term and there is freedom to add more general $q\bar{q}$ contributions,
which can be an argument, in the following, to readjust its normalization. 
Because such contribution ``exist" only in the $-\xi < x < \xi$ region (i.e. they vanish at $\xi=0$), 
they are not sensitive to the relation linking GPDs to quark densities at $\xi$=0.
Furthermore, depending on their parity in $x$, these contributions 
may or may not contribute to the sum rules linking the GPDs to form factors (FFs).
For instance, the D-term is odd in $x$ and doesn't contribute to the form factor 
sum rule.
One sees, in a very general way, that there can be contributions to the 
GPDs which completely escape any normalization or constraint. 

The solid curve on Fig.~\ref{fig:w} shows the results of a calculation with such a 
D-term which has been \underline{renormalized} in order to fit the data.
It is of course not very satisfying to add a term with unconstrained
normalization to the GPDs parametrization. The purpose here
is simply to illustrate that such a term provides just the right $W$ dependence
and is physically motivated in the sense that it is meant to parametrize
$q\bar{q}$ contributions, which are definitely part of the GPD concept. Now, the question 
of course remains
whether one can justify such a strong contribution (the normalization of the D-term 
such as given in Ref~\cite{weiss} barely changes the dashed curve on Fig.~\ref{fig:w}).

\begin{figure}[hbtf]
\begin{center}
\psfig{file=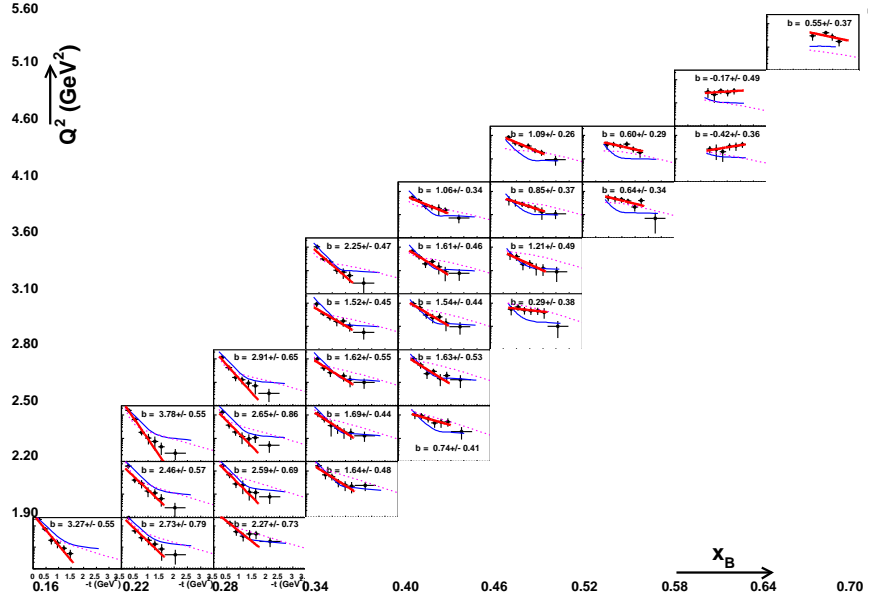,width=4.5in}
\end{center}
\caption{\underline{PRELIMINARY} distribution $dN/dt$ of events for the
$\gamma_L p\to p\rho^0$ process from the CLAS E99-105 experiment. 
Units are arbitrary on the $y$ axis. All $x$ and $y$ axis have
the same scales in each plot. The thick solid line
is an exponential fit $Ae^{bt}$ to the data yielding the $b$ slope parameters.
The thin solid line is the result of a GPD calculation including a 
(``renormalized") D-term-like contribution.
The dotted line is the result of the Regge JML calculation.}
\label{fig:c1}
\end{figure}

The CLAS experiment E99-105, which is in the final stage of analysis, will
soon bring more than 25 new ($Q^2$,$x_B$) data points to this low $W$ domain, along
with differential cross sections. For instance, as a snapshot,
Fig.~\ref{fig:c1} shows the \underline{PRELIMINARY} distribution $dN/dt$ of events for the
$\gamma_L p\to p\rho^0$ process. Because the analysis is still in progress, the normalization 
is arbitrary but the distributions are, for the most part, corrected for the acceptance of CLAS.
One readily sees the large ($Q^2$,$x_B$) domain spanned by these data. What is also obvious
is that the slope of these $t$ distributions vary with the kinematics and, in particular,
increases with decreasing $x_B$. In the framework of GPDs, this is a feature which is
expected because the $t$ dependence of the GPDs can be related~\cite{Burkardt:2000,Diehl:2002he,pire}, 
via a Fourier transform, to the transverse spatial distribution of the partons in the nucleon.
A general ``3-dimensional" image of the nucleon is then that ``low $x$" sea quarks (which
can be associated to the ``pion cloud") sit at the periphery of the nucleon (corresponding to
large impact parameters and therefore large $t$ slopes) while ``large $x$" valence quarks
sit in the core of the nucleon (corresponding to low impact parameters and therefore low $t$ 
slopes). The evolution with $x_B$ of the $t$ slopes of Fig.~\ref{fig:c1} is in agreement
with this image.

Finally, if a D-term like contribution is responsible for most of the cross section
in this low $W$ domain, it remains the ``technical" problem to reconcile a varying
$t$ slope (with energy) with a D-term as the polynomiality rule imposes that the $t$ dependence
of the D-term be factorized (independent of $x$ and $\xi$). We propose here a simple 
ansatz which generalizes the D-term~:

\begin{equation}
GPD^\prime(x,\xi,t)=\xi \int d\alpha d\beta \delta(x-\beta-\xi\alpha)DD^\prime(\alpha,\beta,t)
\end{equation}
with
\begin{eqnarray}
&&DD^\prime(\alpha,\beta,t)=\alpha h(0,\alpha) \frac{b^\prime \mid t\mid}{\mid\beta\mid^{b^\prime t+1}}\nonumber\\
\label{eq:dd2}
\end{eqnarray}
which has the nice feature of recovering the D-term in the forward limit since~:
\begin{equation}
\lim_{\mid t\mid \rightarrow 0} \frac{b^\prime \mid t\mid}{\mid\beta\mid^{b^\prime t+1}} = \delta (\beta)
\end{equation}

\noindent The $\alpha$ factor allows us to respect the polynomiality rule.

The thin solid line in Fig.~\ref{fig:c1} shows the result of the calculation
with such an ansatz and illustrates the variation of the slope with $x_B$
(the $t$ slope is, additionally, also non-constant in this calculation). The dashed 
line is the JML calculation (which uses ``saturating" Regge trajectories) and the thick
solid line is a simple exponential fit to the data, yielding the $b$ slope parameters
which are on the figure.

Let us note the almost flat $t$ slopes at large $x_B$.
Such flat $t$ slopes are basically impossible to reproduce with the
standard double distributions contributions to the $H$ and $E$ GPDs. Indeed,
their $t$-dependence is strongly constrained by the sum rule linking
the GPDs to the FFs. If one is indeed sensitive to GPDs in this low $W$ regime, 
the very flat $t$ slopes observed can only arise from GPD contributions
not constrained by the FF sum rule. 

Let us finally mention that, at large $x_B$, the minimum value of $t$ that is kinematically accessible 
is large (i.e. $\approx$ 1.6 GeV$^{-2}$ at $< x_B >=$ 0.67). A possible explanation
of the disagreement between the ``base" (i.e. without additional D-term-like $t$-channel meson
exchange contribution) GPD model, could simply be that, at such large $t$ values,
higher twists contributions can be extremely important.

Only additional and precise data for the $\gamma_L p\to p\rho^0$ process
will allow us to distinguish between the two hypothesis raised above : GPDs or 
not GPDs ?

With an important new set of data soon to be released, the E99-105
experiment using CLAS will already allow us to study in a more detailed manner
the mechanims at play in this reaction. In the longer term, the
upgrade of JLab to 12 GeV, will permit a precise mapping of the $Q^2, x_B$ and
$t$ dependences over a broad phase space. This, along
with the measurement of new observables such as the transversely
polarized target asymmetry, should bring a more definite answer as to the role or not of GPDs 
in the valence region for exclusive $\rho^0$, and more generally, other meson production.

\vskip .5cm

The authors thank M. Vanderhaeghen, M. Diehl, P. Kroll, D. Mueller,
B. Pire and M. Polyakov for useful discussions.

\bibliographystyle{ws-procs9x6}


\end{document}